\documentclass[a4paper,12pt]{article}
\usepackage[margin=20mm]{geometry}
\usepackage{mathptmx,amsmath,amssymb,amsthm,xcolor,url,hyperref}
\newtheorem*{lemma}{Lemma}
\newtheorem{theorem}{Theorem}
\theoremstyle{remark}
\newtheorem*{remark}{Remark}
\newtheorem*{remarks}{Remarks}

\DeclareMathOperator{\uhr}{\mskip -2 mu\upharpoonright\mskip -1 mu}
\DeclareMathOperator{\KP}{\mathrm{K}\mskip 1 mu}
\DeclareMathOperator{\KM}{\mathrm{KM}\mskip 1 mu}

\let\ge=\geqslant
\let\le=\leqslant

\begin{document}
\title {The Kraft--Barmpalias--Lewis-Pye lemma revisited}
\author{Alexander Shen\thanks{LIRMM, Univ Montpellier, CNRS, Montpellier, France. Supported by ANR-21-CE48-0023 FLITTLA grant.}}
\date{}
\maketitle

\begin{abstract}
This note provides a simplified exposition of the proof of hierarchical Kraft lemma proven by Barmpalias and Lewis-Pye~\cite{bl} and its consequences for the oracle use in the Kučera--Gács theorem (saying that every sequence is Turing reducible to a random one).
\end{abstract}

\section{Kraft's lemma and its online version}

The following statement from coding theory is sometimes called \emph{Kraft's lemma}\footnote{More precisely, the statement is that the \emph{Kraft inequality} mentioned here is necessary and sufficient for the existence of a prefix-free code, see, e.g., \cite[Theorem 3.2.1]{gallager}.}:
\begin{quote}
\emph{for every $n$ integers $l_1,\ldots,l_n\ge 1$ such that $\sum_i 2^{-l_i}\le 1$, there exist binary strings $x_1,\ldots,x_n$ of lengths $l_1,\ldots,l_n$ that form a prefix-free code.}
\end{quote}
The prefix-free requirement means that strings $x_1,\ldots,x_n$ are
incomparable: none of them is a prefix of another one.

It is convenient to identify strings with \emph{aligned intervals} inside $[0,1]$: let $0$ be the left half, $01$ be the second quarter (i.e., $[\frac{1}{4},\frac{1}{2}]$), etc. Formally, a string $x$ corresponds to the interval that contains numbers whose binary representations start with $x$. Then the statement can be reformulated in terms of space allocation. Each $l_i$ is interpreted as the request to allocate an aligned interval inside $[0,1]$ of length  $2^{-l_i}$ in such a way that all intervals are disjoint. This shows immediately that the condition $\sum 2^{-l_i}\le 1$ is necessary for the existence of the prefix code (the total space is bounded). To prove that Kraft inequality is sufficient, we may allocate the intervals in the order of decreasing length (=increasing $l_i$), from left to right. The decreasing length condition guarantees correct alignment. 

However, this allocation strategy needs to know the entire list $l_1,\ldots,l_n$ in advance. A simple change in the allocation strategy makes it \emph{on-line} (getting the next $l_i$, we choose the next $x_i$ and this choice is final). For that, we keep at every moment the representation of the free space as a \emph{union of disjoint aligned intervals of different sizes}. Initially we have one interval of size $1$. When a new $l_i$ arrives, we look for an interval of size $2^{-l_i}$ in the free space list. If there is one, we allocate it (and delete it from the list). If not, we take the minimal larger interval in the free space list, and split it into halves, then one half into two halves, etc., until we get two intervals of size $2^{-l_i}$. One of those intervals is allocated, and all other new parts (including the second interval of size $2^{-l_i}$) are added to the free list.  The minimality guarantees that there are no intervals of that size already in the list. There is only one remaining question: \emph{why the free list contains at least one interval of size at least~$2^{-l_i}$}? If not, all free intervals are strictly smaller than $2^{-l_i}$ and have different sizes that are powers of $2$, so the sum of their lengths is less than $2^{-l_i}$, and that contradicts Kraft's inequality (note that the free space is $1$ minus the total length of already allocated intervals).

This algorithm works for infinite sequences as well, so we get a corollary: 
\begin{quote}
\emph{For every computable sequence of natural numbers $l_i\ge 1$ such that $\sum_i 2^{-l_i} \le 1$,  there exists a computable sequence of incomparable strings $x_i$ of lengths $l_i$.}
\end{quote} 
This result was used by Chaitin to prove the properties of prefix complexity and appears in his paper with the proof presented above (ascribed to N.~Pippenger), see \cite[Theorem 3.2, p.~333]{chaitin}. Now it is often called the \emph{Kraft--Chaitin lemma}. 

Later (also for algorithmic information theory purposes) George Barmpalias and Andrew Lewis-Pye generalized this statement to the case of hierarchical requests~\cite{bl}. In the rest of this note we try to provide an easy-to-read exposition of this result (based on the discussion at the Kolmogorov seminar on complexity; the metaphor of reselling the space was suggested by Bruno Bauwens).

\section{Kraft--Barmpalias--Lewis-Pye lemma}

In the generalized version of Kraft's lemma, formulated and proven by Barmpalias and Lewis-Pye~\cite{bl}, the requests (still being labeled by natural numbers $l_i$) are structured hierarchically. When a new request arrives, it may be declared as a \emph{son} of one of the previous requests. This means that the interval allocated for it should be a part of the father's intervals (instead of being disjoint with all previous intervals). Later this son may get his own sons, etc. 

In other words, requests now form a tree. We add a dummy root for this tree; it will become a father of all requests that had no father. Those requests (of level~$1$, sons of the root) should get disjoint intervals. The requests of level $2$ have fathers of level $1$ (that appeared earlier), etc. The tree grows when a new request arrives: a new leaf is attached to one of existing vertices (the new request becomes a son of some existing request, or a son of a root, if it had no father). Every tree vertex, including the dummy root, may become the father of a new request. 

Formally, each request consists of the natural number $l_i\ge 1$ (its label) and the reference to one of the previous requests or the dummy (root) one.

Now we have to say more precisely what kind of objects should be constructed to satisfy these requests. Let us note first that the space allocation would be simple if the total space requested by all the sons of a vertex never exceeded the space requested by the vertex itself. Then we could use Kraft--Chaitin's allocation process (as described above) at every vertex: each vertex $v$ would take care of the requests of all its sons and give them space inside its own space. The only difference is that instead of a unit interval each vertex gets an interval of some size and the requests from its sons do not exceed that size in total. Note that the requests from the grandsons will be fulfilled by their fathers inside the space allocated to them, so $v$ will not need to worry about them.

We have a quite different setting: we do \emph{not} require that the sum of the requests for sons of some vertex is bounded by the request for the vertex itself. Let us explain the changes needed to adapt the Kraft lemma to this situation.

Recall that the requests form a  growing tree, and every request has a non-negative integer label $l_i$ that means that an aligned interval of size $2^{-l_i}$ is requested. The labels $l_i$ may be arbitrary: for example, a vertex $v$ can have a son $w$ whose request is bigger than the request for $v$; or $v$ can have many sons with total requested size bigger than the request for $v$.  The only restriction for the labels is that the total size $\sum 2^{-l_i}$ is bounded by $1$. Note that this sum includes, for example, both the requests from a vertex $v$ and its son $w$, even if the size of the $w$-request is small and it can be fulfilled inside $v$-space. So this condition is much stronger than necessary for the case we discussed (when sons' requests fit into the father's one).

We make the following changes for the hierarchical version of the lemma:
\begin{itemize}

\item The allocation process is more complex. Initially for a request with label $l_i$ an aligned interval of size $2^{-l_i}$ is allocated. But later more aligned intervals could be allocated to the same request (to the same vertex of the requests' tree). \emph{All these intervals should be of size $2^{-l_i}$ or bigger}. Additional intervals can be allocated at any stage, so the space allocated to a vertex is a (growing) list of disjoint intervals of size at least $2^{-l_i}$ each.
 
\item We have two requirements for allocations. The first says that \emph{the son's space is always inside the father's space}: each interval allocated to the son is a part of some interval allocated to the father.

\item The second requirement says that \emph{brothers have disjoint space}: if $w$ and $w'$ are sons of some vertex $v$, then intervals allocated to $w$ should be disjoint with intervals allocated to $w'$. 
\end{itemize}

Note that for the one-layer tree (root and its sons) we get essentially the statement of the original Kraft lemma, because additional allocated intervals are not helpful in any way.

In other words, the generalized allocation process goes as follows: a new request with label $l_i$ arrives (thus extending the tree); then a new space (an interval\footnote{Or even several intervals of size at least $2^{-l_i}$, though in our construction this would never happen.} of size $2^{-l_i}$) is allocated for this request and some new intervals may be added to the space allocated to other vertices (in fact, this happens only for the tree ancestors of the new request) in such a way that all the conditions mentioned above are satisfied. 

\begin{lemma}[Barmpalias--Lewis-Pye]
A space allocation algorithm that guarantees these properties \textup(assuming that $\sum_i2^{-l_i}\le1$\textup) exists.
\end{lemma}

\section{Proof of the Kraft--Barmpalias--Lewis-Pye lemma}

The allocation algorithm works in a hierarchical way. In the root vertex we have a Kraft space allocator that works as described above. Each vertex of the first level asks the root for an interval of required size and gets it.  If no vertices of the higher level appear, that is all. Vertices of higher level request space from their fathers (so the root allocator does not deal with their requests directly).

It is useful to describe this process in terms of buying and reselling space for a fixed price (aligned interval of size $2^{-l}$ costs  $2^{-l}$). If there are no hierarchical requests, we are in the situation of standard Kraft--Chaitin lemma: the root allocator sells space inside $[0,1]$ to customers (i.e., requests of level $1$), keeping the information about remaining free space as a list of disjoint intervals. Initially the root allocator has no money, but owns the entire interval (the free list contains one interval of size $1$). Gradually the amount of space in its possession decreases, and the amount of money increases. The sum (space + money) always remains the same ($1$). The root allocator never runs out of space because $\sum 2^{-l_i}\le1$.

In this case (no hierarchical requests) no reselling is happening, and each customer comes to the root allocator only once. Both things change when hierarchical requests appear. As we mentioned, a request (a node of the requests' tree) of level greater than~$1$ never talks directly with the root allocator. Instead, it speaks only with its father, who may resell some space bought earlier, or --- if needed ---may buy more space from its own father and resell all or part of this space to its son. In this scheme a node with label $l$ can sequentially request several intervals from its father, but none of these intervals should be smaller than $2^{-l}$ (the size of the original request). Therefore, to satisfy small requests of its sons, a node should aggregate their requests, buying the space in big chunks and reselling it in smaller ones. 

Let us note that 
\begin{itemize}
\item
each seller does not make any difference between new and old customers: all requests are processed in the same way; 
\item
a vertex that does not have enough space requests additional space from its father, so one request may trigger a chain of actions (that may propagate to the root).
\end{itemize}

To finish the proof, we should describe the aggregation algorithm and prove its correctness (this means that no vertex will run out of money or space).

Initially, a request with label $l$ has $2^{-l}$ units of money. It uses this money to buy an (aligned) interval of size $2^{-l}$ from its father. After that it has no money, only the space. Then it starts reselling the space to its children (when/if  they arrive). In this process it may need to buy additional amount of space from the father. Here is the algorithm for the vertex:

\begin{itemize}
\item Keep the information about the space you own as the list of disjoint aligned intervals of different sizes; note that the size of this space plus the amount of money you have is always $2^{-l}$, where $l$ is your label.
\item If an interval of some size is requested, and an interval of exactly this size exists in the list, then sell this interval (and delete it from the list). The space reserve decreases and the money reserve increases (by the same value, the length of the resold interval).
\item If there is no interval of the requested size in the free list, but there is a bigger free interval, then split this bigger interval into two halves, split one half in two halves, etc., until two intervals of the requested size appear. Sell one of them, and keep the other (and all bigger new intervals) in the free list. Again, the space reserve decreases, and the money reserve increases.
\item It may happen also that you get a request of size $2^{-l}$ or bigger. In this case buy an interval of the requested size from your father and immediately resell it to your son. (Free space and the amount of money remain the same.)
\item Finally, it may happen that an interval of size smaller that $2^{-l}$ is requested but all free intervals are smaller than the requested one. Since free intervals are of different sizes, this implies that the total amount of the free space is smaller than the requested interval. In this case you are low in space, but high in money: the amount that is missing for buying a new interval of size $2^{-l}$ (this missing amount is equal to the size of free space) is smaller than the size (=price) of the requested interval.  Use the money reserves plus the customer payment to get a new interval of size $2^{-l}$ and split is as before, then give its part to the customer, and put the rest in the free list. (Since the existing free intervals were smaller than the request, the list again will consist of intervals of different sizes.)
\end{itemize} 

In this scheme no cash is injected except for the $2^{-l_i}$ amounts initially given to the requests, so the root allocator will never run out of space  (since $\sum 2^{-l_i}\le 1$). Note that the description of the allocation algorithm does not refer to money at all; all this accounting (similar to what is done sometimes for amortized analysis) is needed only to prove that the root allocator will never run out of space.

The Kraft--Barmpalias--Lewis-Pye lemma is proven.

\begin{remark}
We considered the case of binary alphabet. If we have $m$ letters, we get a tree with branching factor $m$, and the Kraft inequality has the form $\sum m^{-l_i}\le 1$. Both the original Kraft--Chaitin argument and the proof of Kraft--Barmpalias--Lewis-Pye lemma can be easily adapted to this case. Now the invariant is that the list of free intervals may contain at most $m-1$ copies of the intervals of the same size. (For $m=2$ we get the previous requirement: all intervals are different.) In this way, the numbers of intervals of each size correspond to the digits in the $m$-ary representation of the amount of the free space. The allocation algorithm remains essentially the same: if there is an interval of the required size, allocate it; if not, take the minimal bigger free interval and split it into $m$ pieces, then do the same for one of the pieces, etc. This process corresponds to subtracting $1$ from $m$-ary number $100\ldots0$: we get $(m-1)$ new free intervals of all intermediate sizes.
\end{remark}

\section{Positive result: efficient coding}

The Kraft lemma has a natural interpretation in terms of coding; it allows us to construct a prefix code for $k$ letters with codewords of lengths $l_1,\ldots, l_k$ assuming that $\sum_{i=1}^k 2^{-l_i}\le 1$. The Kraft--Chaitin lemma extends this result to countably many letters. The Kraft--Barmpalias--Lewis-Pye provides \emph{hierarchical coding}: we require that codes of some letters are extensions of the code of others letters, so the tree structure of letters should be preserved in the tree of codes. Using the compactness argument, we may get a similar conclusion for \emph{infinite} branches. Note that the Kraft--Barmpalias--Lewis-Pye lemma is valid both for finite and infinite sequences of requests. In the latter case the requests' tree grows when new requests arrive, and we can consider the limit (infinite) tree  of requests which includes all the requests.

\begin{lemma}
Assume that the tree of requests has an infinite branch $r_1,r_2,\ldots$, where $r_1$ is the request of the first level, and $r_{i+1}$ is a son of $r_i$. Then there exists a sequence of strings $x_1,x_2,\ldots$ such that $x_i$ is a prefix of $x_{i+1}$ for all $i$, and every $x_i$ is one of the codes of $r_i$ obtained by Barmpalias--Lewis-Pye construction (and, therefore, the length of $x_i$ does not exceed the label of the request $r_i$).
\end{lemma}

Note that several strings (intervals) may correspond to the same request $r_i$, and we claim that one can choose one of them ($x_i$) for every $i$ in such a way that $x_i$ is a prefix of $x_{i+1}$. This choice is not effective, though.

\begin{proof}
Let $x_i$ be arbitrary string that is allocated to $r_i$ during the construction. Then the corresponding interval is inside the space allocated to $r_{i-1}$, so $x_i$ has some prefix $x_{i-1}$ that earlier was allocated to $r_{i-1}$. Then we can find some prefix $x_{i-2}$ of $x_{i-1}$ allocated to $r_{i-2}$, etc. 

The only problem is that for different $i$ we get different sequences $x_1,\ldots,x_i$, so we do not get directly an infinite sequence $x_1,x_2,\ldots$ of strings allocated to $r_1,r_2,\ldots$. We need to use compactness argument (K\"{o}nig's lemma). Note that every request $r_i$ has only finitely many strings allocated to it (at most $2^{-l}$ if the label is $l$). So some $x_1$ appears for infinitely many $i$. Choosing this $x_1$ and retaining only the values of $i$ when this $x_1$ is used, we then choose some $x_2$ that appears infinitely many times, etc. 
\end{proof}

Note that this argument does not provide a \emph{computable} sequence of $x_i$ even if both the sequence of requests and the branch $r_1,r_2,\ldots$ are computable.

This lemma implies the following result (which was one of the main goals of~\cite{bl})

\begin{theorem}\label{th:reduction}
Let $K$ be a total computable function on binary strings such that $\sum_x 2^{-K(x)}\le 1$. Then there exists an oracle machine $M$ with the following property: for every bit sequence $\alpha$ there exists a bit sequence $\beta$ such that $M$ computes $\alpha$ with oracle $\beta$, and the oracle use when computing the prefix $\alpha \uhr n$ is at most $K(\alpha\uhr n)$.
\end{theorem}

The oracle machine has an \emph{input tape} where an infinite sequence of zeros and ones (the \emph{oracle}) is written. The machine reads the input tape bit by bit, while performing some other computations, and writes the output bit sequence (one bit at a time). For a given oracle $\beta$ the output sequence $\alpha$ may be finite or infinite; for every prefix $\alpha\uhr n$ of $\alpha$ we consider the number of input bits read up to the moment when $n$ output bits were produces, and this number is called the \emph{oracle use}.

\begin{proof}
Let us consider strings $0$ and $1$ as two requests of the first level with labels $K(0)$ and $K(1)$; then $00$ and $01$ are requests of level $2$ that are sons of the request $0$ and have labels $K(00)$ and $K(01)$; in the same way $10$ and $11$ are sons of $1$, etc. Applying Kraft--Barmpalias--Lewis-Pye lemma, we get a computable sequence of allocations.

Let us first look on the strings allocated to the requests of the first level ($0$ and $1$). They form an enumerable prefix-free set of strings that consists of two disjoint parts: codewords for $0$ and codewords for $1$. We need to construct an oracle machine that outputs $0$ if the oracle has a prefix of the first type, and outputs $1$ if the oracle has a prefix of the second type.  It would be trivial without the additional requierements (just read the oracle bits and enumerate these two parts in parallel), but we want that the machine \emph{does not read any bits after the codeword}. To satisfy this additional requirement, we may delay the reading the next bit \emph{until some codeword appears that is a proper extension of an already read prefix $z$ of the oracle}. If this never happens, either $z$ is a codeword itself (and we will find this out at some point, and produce the output bit as required), or $z$ is not a prefix of any codeword (then we produce no output, and this is the right behavior).  If this happens at some point, then we know that (because of prefix-free requirement) $z$ is not a codeword, so we can safely read the next bit, and continue in the same way. (Formally we maintain the following invariant: any proper prefix of the already read part of oracle is not a codeword, see~\cite[Theorem 50, p.~86]{suv} for the details.)

After a codeword for $0$ or $1$ is read, we perform the same operation for the next bit. For example, if the codeword for $0$ is read, we are looking for its extensions that are codewords for $00$ and $01$ in the same way. Theorem is proven.
\end{proof}

\begin{remarks}
\leavevmode\par
\begin{enumerate}
\item
Note that the oracle use is monotone, so the bound implies that the oracle use is at most $\min_{i\ge n} K(\alpha\uhr i)$.

\item
The same argument works if $K$ is a partial computable function (with natural values) defined on a subtree of the full binary tree: a new vertex to the requests tree is added when the value of $K$ on the corresponding string is computed. Moreover, if $K$ is a computable finction with an arbitrary domain, we may restrict $K$ to the maximal subtree inside the domain of $K$ (by checking whether $K$ is defined on all prefixes). In this way we get a similar statement where $K$ is an arbitrary partial computable function and we additionally assume that $K(x)$ is defined for all prefixes $x$ of a given sequence $\alpha$ (that can be now finite or infinite).
\end{enumerate}

\end{remarks}

\section{Contaminated space and reduction to random sequences}

A stronger version of Theorem~\ref{th:reduction} guarantees that the sequence $\beta$ is Martin-L\"of random\footnote{From now on we assume that the reader is familiar with algorithmic randomness and Kolmogorov complexity theory; all the needed notions and results can be found, e.g., in~\cite{suv}.}. According to the definition, the set of all non-random sequences is contained in an effectively open set of arbitrarily small measure. Now we assume that $\sum_x 2^{-K(x)}<1$ (strict inequality) and take an effectively open set $U$ whose measure is smaller than the gap between $\sum_x 2^{-K(x)}$ and $1$. Then we use the same argument with the following stronger version of Barmpalias--Lewis-Pye lemma.

Again, we consider a sequence of hierarchical requests. In addition we have a parallel process that enumerates a set of aligned intervals that are considered as ``contaminated''. So at every moment we have an clopen contaminated subset of the unit interval that increases with time. (Only this subset matters; we do not care how this subset is split into a union of intervals.) We assume that that the contaminated part remains small; namely, we assume that the total size of all the requests \emph{plus the size of the contaminated part} never exceeds $1$.

\begin{lemma}
In this case one can arrange the allocation process for all the requests with the following additional requirement: when an interval is allocated, it is not completely contaminated, i.e., is not covered by the part of the contaminated space known at the moment of the allocation.
\end{lemma}

\begin{remark}
Note that the statement of the lemma does not prevent the following cases:
\begin{itemize}
\item part of the interval just allocated is contaminated;
\item at some later stage the entire interval will get contaminated.
\end{itemize}
\end{remark}

\begin{proof}
Let us add to our picture the following insurance service: if an allocated interval (obtained from the father node) turns out to be entirely in the contaminated space at the moment of its allocation, the insurance reimburses you for the amount paid for this interval (i.e., its length), but you cannot use this interval later for reselling. With the insurance money you can then buy another interval of the same length, and if again it turns out to be completely inside the contaminated space, you again get the money back, and then buy one more interval, etc.. This process will stop at some point (since there are only finitely many intervals of the same length).

Note that you cannot get reimbursement later (only at the moment of allocation and only if the allocated interval is completely in the already contaminated zone).

The only thing to check is that the total amount of money paid by all clients during the allocation is at most $1$. This money comes from two sources: the initial money the clients have, and the money paid by the insurance. Note that the insurance service never reimburses the same space twice: if an interval is reimbursed, it is blocked for the future use, and neither its owner or his brothers can use it. So when an interval is reimbursed, it is disjoint with all previously reimbursed intervals.
\end{proof}

Now this lemma can be used to prove the following stronger version of coding theorem:

\begin{theorem}\label{th:reduction-random}
Let $K$ be a partial computable function on binary strings such that $\sum 2^{-K(x)}< 1$. Then there exists an oracle machine $M$ with the following property: for every bit sequence $\alpha$ such that $K(\alpha\uhr n)$ is defined for all $n$, there exists a Martin-L\"of random bit sequence $\beta$ such that $M$ computes $\alpha$ with oracle $\beta$, and the oracle use when computing the prefix $\alpha \uhr n$ is at most $K(\alpha\uhr n)$.
\end{theorem}

Note that the inequality is strict here, and the sequence $\beta$ is guaranteed to be random. This is a strong version of Kučera--Gács theorem saying that every sequence is Turing reducible to a random one; see~\cite{bs} about the (rather long) history of improvements on the oracle use bound in this theorem and related results.

\begin{proof}
The proof goes as before, but we start with taking an effectively open set $U$ that covers all non-random sequences and whose measure is small enough, so together with $\sum 2^{-K(x)}$ it does not exceed~$1$. Then we can apply the lemma, using the enumeration of this effectively open set as a generation of contaminated space, and get in the same way the sequence $\beta$ that computes~$\alpha$. We need to show only that $\beta$ (the limit sequence) does not belong to $U$. Assume that $\beta$ belongs to $U$; then some prefix $b$ of $\beta$ has the corresponding interval completely covered by $U$ (since $U$ is open). Then at some point (as the compactness argument shows) the interval corresponding to $b$ will be completely contaminated.  Starting from this moment, all the smaller intervals are also fully contaminated and cannot be allocated. On the other hand, only finitely many prefixes of $\beta$ are allocated before this moment, so we get a contradiction.
\end{proof}

\section{Prefix complexity and oracle use}

We assumed that the function $K$ (the upper bound for the oracle use) was computable; the natural question is whether this result can be extended to upper semicomputable functions, or, equivalently, whether a bound with prefix complexity is valid. 

This question is mentioned as open in the paper of Barmpalias and Lewis-Pye~\cite[p.~4484, Conjecture]{bl2}. Here is the exact statement. Let $\KP$ be the prefix complexity function (not to mix with italic $K$ that appeared earlier and was a computable upper bound for $\KP$).
\begin{quote}
Let $\alpha$ be an arbitrary sequence. Can we always construct a sequence $\beta$ and the oracle machine $M$ such that $\alpha$ is computed by $M$ with oracle $\beta$ and the oracle use for $n$-bit prefix $\alpha\uhr n$ of $\alpha$ is bounded by $\min_{i\ge n} \KP(\alpha\uhr i)+ c$ for some $c$ and all  $n$?
\end{quote}

A weaker result\footnote{It would be interesting to find an easy proof of this result by adapting the arguments explained above.} appears in the same paper as Theorem I.5: it replaces $K(\alpha\uhr i)$ by $[K(\alpha\uhr i)+\log_2 i]$. 

One can consider a stronger conjecture where the bound is replaced by $\min \KP(y)$ over all $y$ that are extensions of $\alpha\uhr n$ (and not only for prefixes of $\alpha$). For this stronger conjecture the answer is negative. The counterexample, a sequence $\alpha$ that does not have this property, can be (as proven by Mikhail Raskin, who kindly permitted to include his argument) constructed diagonally. At each step we extend the existing prefix $a$ of $\alpha$ to some longer $a'$ preventing some machine $M$ from satisfying the requirement with some constant~$c$. (There are countably many pairs $M,c$, so we can diagonalize against all of them.)

So let us assume that $a$, $M$ and $c$ are fixed. We want to find some $b$  that extends $a$ and has the following property: machine $M$ cannot compute any infinite extension of $b$ with required bound for the oracle use. To achieve this goal, we select some computable infinite sequence that starts with $a$ (for example, we can write all zeros after $a$) and let $a'$ be a long prefix of this sequence whose prefix complexity is negligible compared to length. (To specify the prefix of a computable sequence, we use finitely many bits to specify the program, and the remaining bits are used to specify its length that can be very large compared to the complexity.)

Now $\KP(a')$ is fixed, and we may try all strings of size $\KP(a')+c$ as oracles for machine $M$. Each of them computes some sequence (finite or infinite), and we compare all these sequences with~$a$ and~$a'$. Several cases are possible:
\begin{itemize}
\item Some of these sequences do not go through $a$ at all.
\item Some of them start with $a$ and then stop or deviate from $a'$.
\item Finally, some other could reach $a'$.
\end{itemize}
In any case, since there are many prefixes between $a$ and $a'$, one can find some $a''$ with the following property: any program that reaches $a''$, makes one more step in the direction of $a'$. Then finally we let $b$ be the extension of $a''$ that deviates from the path to $a'$.

The chosen $b$ has the required property. Let $\beta$ be any infinite extension of $b$. Assume that $M$ computes $\beta$ with some oracle $\alpha$ with the bounded oracle use. Since $a''$ is a prefix of $\beta$ and at the same time prefix of $a'$, the oracle use for $a''$ should be at most $\KP(a')+c$. But all oracles that compute $a''$ with this oracle use, compute also the next bit of $a'$ and therefore are unsuitable for $b$.

\begin{remark}
A more accurate accounting disproves a weaker conjecture where $\min\KP(y)$ over all extensions of $\alpha\uhr n$ is replaced by $\min[\KP(y)+0.99\log|y|]$ over the same extensions. Indeed, we need that the number of possible oracle prefixes of size $\KP(a') + 0.99\log |a'|+c$ is smaller than the difference between the lengths of $a'$ and $a$ (so we can find some $a''$ with required properties). The length of $a$ is fixed, so we need that $2^{[\KP(a')+0.99\log |a'|+c]}\ll a'$, and this is possible since $\KP(a')+c$ can be small compared to $0.01\log |a'|`$.

Note that this example shows also that one cannot always find $\beta$ that computes $\alpha$ with oracle use $\KM(\alpha\uhr n)$, where $\KM$ stands for monotone complexity. Indeed, since $\KM(z)\le \KP(z)$ for all $z$ and $\KM$ is monotone, we have $\KM(x)\le \min \KP(z)$ when the minimum is taken over all extensions $z$ of $x$. (This question is natural since $\KM(\alpha\uhr n)$ is an obvious lower bound for the oracle use.)

\end{remark}

\section*{Acknowledgements}

This paper is based on the discussions with George Barmpalias, Bruno Bauwens, Laurent Bienvenu, Michael Raskin, Mikhail Vyalyi and other participants of \emph{Kolmogorov seminar} on description complexity. I thank the (anonymous) reviewers of the CCR2023 conference for their comments.

\end{document}